\documentclass[11pt, a4paper]{article}

\usepackage{color}
\usepackage[dvipsnames]{xcolor}
\usepackage{graphicx}
\usepackage{amssymb}
\usepackage{amsmath}
\usepackage{braket}
\usepackage{fontawesome}
\usepackage{soul}
\usepackage[english]{babel}
\usepackage[T1]{fontenc}
\usepackage[noadjust]{cite}
\usepackage{hyperref} 

\usepackage{bm}

\usepackage[utf8]{inputenc}
\usepackage{authblk}

\usepackage{qcircuit}

\newcommand\eeq{\end{equation}}
\newcommand\beq{\begin{equation}}
\newcommand\eea{\end{eqnarray}}
\newcommand\bea{\begin{eqnarray}}

\usepackage{listings}
\usepackage{xcolor}
\definecolor{codegreen}{rgb}{0,0.6,0}
\definecolor{codegray}{rgb}{0.5,0.5,0.5}
\definecolor{codepurple}{rgb}{0.58,0,0.82}
\definecolor{backcolour}{rgb}{0.95,0.95,0.92}

\lstdefinestyle{mystyle}{
    backgroundcolor=\color{backcolour},   
    commentstyle=\color{codegreen},
    keywordstyle=\color{magenta},
    numberstyle=\tiny\color{codegray},
    stringstyle=\color{codepurple},
    basicstyle=\ttfamily\footnotesize,
    breakatwhitespace=false,         
    breaklines=true,                 
    captionpos=b,                    
    keepspaces=true,                 
    numbers=left,                    
    numbersep=5pt,                  
    showspaces=false,                
    showstringspaces=false,
    showtabs=false,                  
    tabsize=2
}

\begin{document}

\linespread{1.1}

\title{
\textcolor{Mulberry}{\bf Simulating Bell inequalities with Qibo}}

\author[1,2]{ {\large\sc Isabella Masina} \thanks{Corresponding author: masina@fe.infn.it}}
\author[3]{ {\large\sc Giuseppe Lo Presti} \thanks{giuseppe.lopresti@cern.ch}}
\author[3,4]{ {\large\sc Matteo Robbiati} \thanks{matteo.robbiati@cern.ch}}
\author[3]{ {\large\sc Michele Grossi} \thanks{michele.grossi@cern.ch}}

\affil[1]{\small\it Dept.\,of Physics and Earth Science, Ferrara University, Via Saragat 1, 44122 Ferrara, Italy }
\affil[2]{\small\it Istituto Nazionale di Fisica Nucleare (INFN), Sez.\,di Ferrara, Via Saragat 1, 44122 Ferrara, Italy }
\affil[3]{\small\it CERN, Europen Organization for Nuclear Research, Geneva, Switzerland}
\affil[4]{\small\it TIF Lab, Dipartimento di Fisica, Universit\`a degli Studi di Milano, Milano, Italy}


\maketitle

\begin{abstract}

We present educational material about Bell inequalities in the context of quantum computing.
In particular, we provide software tools to simulate their violation, together with a guide for the classroom discussion. 
The material is organized in three modules of increasing difficulty, and the relative implementation has been written in Qibo, an open-source software suite to simulate quantum circuits with the ability to interface with quantum hardware. 
The topic of inequalities allows not only to introduce undergraduate or graduate students to crucial theoretical issues in quantum mechanics -- like entanglement, correlations, hidden variables, non-locality --, but also to practically put hands on tools to implement a real simulation, where statistical aspects and noise coming from current quantum chips also come into play.

\end{abstract}

{\flushleft Keywords: Quantum Computing; Bell inequalities; Qubits; Quantum Technologies; Education}

\linespread{1.2}

\vskip 1.cm
\section{Introduction}

A basic understanding of quantum physics is a necessary prerequisite to work in professions related to second generation quantum technologies. 
In the last years we assisted to a multiplication of courses on quantum computing. Professors involved in such courses would clearly benefit from having educational tools at their disposal. 
Various educational materials and tools have indeed been released recently, allowing to explore quantum mechanical effects: a reference portal for the quantum education community is QTEdu \cite{web-QTEdu}, launched as part of the Quantum Flagship research and innovation programme funded by the EU \cite{Riedel_2019}. Interesting approaches to prepare a course on quantum computing can be found {\it e.g.} in \cite{Haghparast:2024cnc} and \cite{Sun2024}. 

In this respect, the topic of Bell inequalities represents a fascinating and instructive case study.
The goal of this contribution is to present an educational tool primarily targeted to graduate students, useful for the sake of introducing concepts related to Bell inequalities and simulating their violation by quantum mechanical systems.  
This offers the opportunity of dealing with theoretical arguments -- like entanglement, spin correlations, hidden variables, non-locality -- and discussing how they can be visualized, simulated and implemented in a real situation, 
where also statistical aspects and noise come into play.  
The tool proposed in this work consists of a set of software simulations~\cite{simcode} based on Qibo~\cite{qibo_paper}, an open-source quantum computing framework.
Other well known software frameworks for quantum computing developments include Qiskit~\cite{qiskit}, Cirq~\cite{cirq}, Qulacs~\cite{qulacs}, and PennyLane~\cite{pennylane}.

Bell inequalities are by now a popular topic not only for the scientific community, but also for the general public, due to the assignment of the 2022 Nobel Prize in Physics to A.\,Aspect, J.\,Clauser and A.\,Zeilinger, \textit{for experiments with entangled photons, establishing the violation of Bell inequalities and pioneering quantum information science}.
Interestingly, this topic is increasingly linked to high-energy physics, as quantum effects can also be measured from data extracted at the CERN LHC experiments\cite{grossi2024}. Indeed, the highest-energy observation of entanglement in top-antitop quark events was recently reported in \cite{2024ATLAS}.

As it is well known, there are many versions of Bell inequalities. 
The original version proposed by J.\,Bell in 1964 \cite{Bell:64} derived inspiration from the much debated Einstein Podolski Rosen (EPR) paper \cite{EPR:35,Bohr:35}, 
in particular according to its formulation in terms of two spin-$1/2$ particles given by D.\,Bohm \cite{Bohm:51}.
This inequality inspired many other similar inequalities, more suitable to actual experimental setups. 
This was the case for the so-called CHSH inequality proposed in 1969 by J.\,Clauser, M.\,Horne, A.\,Shimony  and R.\,Holt \cite{CHSH:69}, and its successive variations. 
In 1970, E.\,Wigner elaborated a version of Bell’s inequality that has the advantage of being particularly intuitive \cite{Wign:70}, although not actually convenient for experimental setups; 
the Bell-Wigner inequality was chosen to introduce the topic of Bell inequalities in many quantum mechanics textbooks, as the renowned one by J.J.\,Sakurai \cite{Saku:85} 
and the more recent book by G.\,Fano and S.M.\,Blinder \cite{Fano:17}. 

In 1972, J.\,Clauser and S.\,Freedman carried out the first experimental test of the CHSH-Bell's theorem predictions; 
this was the first experimental observation of a violation of a Bell inequality \cite{FreCla:72}.
In 1974, J.\,Clauser and M.\,Horne \cite{ClaHor:74} showed that a generalization of Bell's theorem provides severe constraints for all local realistic theories of nature (a.k.a. objective local theories); that work introduced the Clauser–Horne inequality as the first fully general experimental requirement set by local realism. For more details, see {\it e.g.} the review by J.\,Clauser and A.\,Shimony \cite{ClaShi:78}, and J.\,Bell's fun analogy with socks \cite{Bell:1980wg}. Bell inequalities were later generalized to the case of $n$ spin-$1/2$ particles by D.\,Mermin \cite{Mermin:1990vxe} (see also \cite{PhysRevA.94.012314, PhysRevA.94.032102}). 

It has to be admitted that CHSH-type inequalities and even the original Bell inequality are mathematically 
and conceptually challenging for students in physics, not to say for a wider audience; 
various tools allowing to simulate violations of CHSH
inequalities have been nevertheless proposed, based on the Qiskit \cite{Qiskit:CHSH} and Qibo~\cite{Qibo:CHSH} open source platforms.
As a matter of fact, the Bell-Wigner inequality is accessible to a wider audience of students than is the case of the CHSH-type inequalities: the required mathematics is trivial and the basic assumption can be visualized in a simple way, as we are also going to discuss. 
At the best of our knowledge, tools for simulating the Bell-Wigner inequality or the original Bell inequality lack. 

In the following, we thus start describing how to simulate the Bell-Wigner inequality, secondly we turn to consider the original Bell inequality, and finally introduce the CHSH-type inequalities. Thanks to the unified notation and the increasing level of difficulty, we hope this contribution will represent a useful educational tool for courses in quantum computing, as well as for outreach events. 

Quantum educational material that complements ours is the construction of a setup for measuring Bell inequalities, as discussed in \cite{Sanz2024}. Related educational material aimed at high-school students, including encoding of polarization and which-path information of a photon, can be found in \cite{Zuccarini2024}.

This paper is organized as follows.
Sec. \ref{sec-Pre} discusses the prerequisites and provides instructions for using the material presented here.
Secs. \ref{sec-BW},
\ref{sec-B} and \ref{sec-CHSH} are three modules devoted respectively to the Bell-Wigner, the original Bell and the CHSH-type inequalities.
In sec.\,\ref{sub-thfr-WB} we review the theoretical framework of the Bell-Wigner inequality and its assumptions; in sec.\,\ref{sub-qc-WB} and sec.\,\ref{sub-qibo-WB} we discuss the related quantum circuit and its implementation in Qibo. 
Secs.\,\ref{sub-thfr-B} and \ref{sub-qibo-B} are devoted to the theory and implementation of the original Bell inequality.
Secs.\,\ref{sub-thfr-CHSH} and \ref{sub-qibo-CHSH} deal with the theory and implementation of CHSH-type inequalities. 
Sec.\,\ref{sec-sn} presents a discussion of aspects related to statistics and noise. 
We conclude in sec.\,\ref{sec-conc}.
App.\,\ref{app-rhombo} is devoted to a visualization of the populations endowed with local hidden variables according to Wigner's argument, and App.\,\ref{app-qibo} provides a short introduction on how Qibo implements state-vector simulation to execute a quantum circuit.
For the sake of completeness, in app.\,\ref{app-Bell} we review the mathematical derivation of the Bell inequality \cite{Bell:64}.


\section{Prerequisites and instructions for use}
\label{sec-Pre}

This activity is primarily intended for Master students in Physics who already had a Bachelor course on Quantum Mechanics. They should already be familiar with the bra-ket notation, spin measurement results, and composition of angular momenta, including spin singlet states.

The material is ideally placed in an introductory course on Quantum Computing as a laboratory activity.
In terms of preliminary knowledge on quantum computing, students should be familiar with unitary quantum gates, Bell states, and qubits manipulation through elementary quantum circuits.

Alternatively, the activity may be proposed to Computer Science students approaching quantum computing, or in topical advanced Schools on Computing: in this case, while students may not grasp the full depth of the theoretical considerations, they can take significant benefit from the laboratory activity in order to explore the behaviour of entangled qubits in Bell states.

The code for the simulations \cite{simcode} is based on Python notebooks and can be executed on a standalone Jupyterlab\footnote{The Jupyterlab software is available for download at \url{https://jupyterlab.org}} setup.
Students should be familiar with installing the required packages in such a software environment, which is often already available in a teaching or academic context\footnote{At CERN, Qibo has been made available in SWAN \cite{SWAN:18}, a Jupyterlab-based service widely used for interactive physics analyses by the research community.}. Alternatively, the packages can be installed in the Google \textit{Colab}\footnote{\url{https://colab.research.google.com}} environment, which is a publicly available environment for data science.
The notebooks include detailed step-by-step instructions and allow for variations of the resulting plots.

The theoretical material may be discussed in up to two hours, whereas the laboratory activity may take about three to four hours. Of course, the module may take more time depending on how much the teacher and the students wish to dive into the details. The material is flexible enough for the teacher to tailor the class according to the students' interest and curiosity.

\section{The Bell-Wigner inequality}
\label{sec-BW}

As a first introductory module, we discuss the Bell-Wigner inequality\,\cite{Wign:70}.
After introducing the theoretical framework, we then discuss how to construct an experiment through a proper quantum circuit. We finally describe a more practical implementation using Qibo. This module allows to set the notation and prepares the ground for the subsequent modules about the Bell and CHSH-type inequalities.

\subsection{Theoretical framework}
\label{sub-thfr-WB}

\begin{figure}[h!]
\vskip .5cm 
 \begin{center}
    \includegraphics[width=12.5cm]{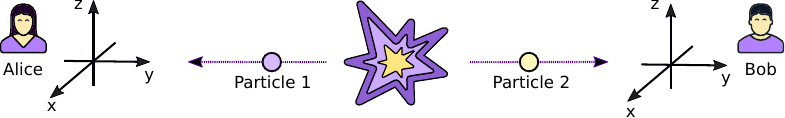} 
    \end{center}
\caption{\baselineskip=15 pt \small 
Experimental setup to study the Bell-Wigner inequality.
}\label{fig-AB}
\vskip .5 cm
\end{figure}

Following the notation and discussion in \cite{Saku:85}, we consider the setup of Bohm's version of EPR\,\cite{Bohm:51}.  The starting point is an entangled state, specifically the Bell state corresponding to the spin singlet combination of two spin-$1/2$ particles, that is 
\beq
|\Psi^- \rangle=\frac{1}{\sqrt{2}}(|+-\rangle-|-+\rangle) \,\, ,
\label{eq-singlet}
\eeq
where the first and second label respectively refer to the sign of the spin component along $\hat z$ of the first and second particle. By construction, 
the orbital angular momentum of the system is vanishing, as it is the total spin (intrinsic angular momentum). Therefore, the total angular momentum of the system, which is a conserved quantity, is vanishing. The experimental setup is shown in fig.\,\ref{fig-AB}.

The system in the final state is assumed to be made of two spin-$1/2$ particles flying apart, say along the $y$ direction, with null relative orbital angular momentum. The latter requirement is important as, due to angular momentum conservation, this implies that the particles still are in a total spin singlet state.
Observers $A$ (Alice) and $B$ (Bob) measure the spin component of the first and second particle, respectively; the measurement is done along some direction, and the result they find is $+1$ or $-1$, in units of $\hbar/2$. The results of $A$ and $B$'s measurements are denoted by $\alpha$ and $\beta$, respectively. 

Consider for instance the case in which $A$ and $B$'s both measure the spin along $\hat z$.
The initial singlet state (\ref{eq-singlet}) has vanishing spin component along $\hat z$ (as in any other direction); this must apply to the final state too. Due to the additivity of the spin components along any given direction, the final state has vanishing total spin component along $\hat z$ if $\alpha+\beta=0$, 
or equivalently $\alpha\,\beta=-1$. 
The observers should thus never find $\alpha\,\beta=1$, as this would represent a violation of angular momentum conservation along $\hat z$. 
Instead, in the case where the observers measure the spin in opposite directions, say $+\hat z$ and $-\hat z$, conservation of angular momentum implies $\alpha-\beta=0$, or equivalently $\alpha\,\beta=1$. In summary, angular momentum conservation
implies that when $A$ and $B$ measure the spin in the same (opposite) direction, they find that the product of their measurements is $\alpha\,\beta=-1(+1)$.
The outcomes $\alpha$ and $\beta$ are thus perfectly correlated for measurements in the same or opposite directions.

All the matter is about the origin of such correlations. On one hand, it would be surprising not to see them, as they account for angular momentum conservation; on the other hand, what is actually surprising is the fact that the observed correlations are found measuring particles that can be very far apart. This calls for an explanation in terms of some fundamental mechanism (see Bell's nice analogy with socks \cite{Bell:1980wg}): is the result of a spin component measurement "predetermined"\,\footnote{According to EPR \cite{EPR:35}, predetermination is related to the existence of an "element of physical reality". A sufficient condition for the latter is: If, "without in any way disturbing a
system, we can predict with certainty (i.e., with
probability equal to unity) the value of a physical
quantity, then there exists an element of physical
reality corresponding to this physical quantity."} before any measurement is done on the particles? Or is the result affected by a possible measurement on the other particle?
Only the first option would be compatible with locality, the second requiring, adopting Einstein's words, a "spooky action at distance". 
Now, in order to formally introduce "predetermination",
one needs a hidden variable theory: a deterministic physical model that seeks to explain the probabilistic nature of quantum mechanics by introducing additional inaccessible variables. 
A local hidden variable theory is a hidden variable theory that satisfies the principle of locality, stating that an object is influenced only by its immediate surroundings, in contrast to the concept of instantaneous, "non-local" action at a distance, which is embraced by the standard interpretation of quantum mechanics, based on the "collapse" of the wave function.

\begin{figure}[h!]
\vskip .5cm 
 \begin{center}
    \includegraphics[width=12.5cm]{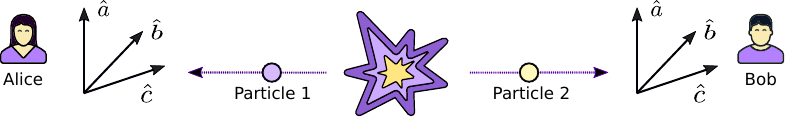} 
    \end{center}
\caption{\baselineskip=15 pt \small 
Three directions for the experimental setup studying the Bell-Wigner inequality.
}\label{fig-abc}
\vskip .5 cm
\end{figure}

A convenient setup to study the correlations is the following: 
each observer measures the spin along one among three possible directions, denoted by unit vectors $\hat a, \hat b$ and $\hat c$, as in fig. \ref{fig-abc}.

Following Wigner\,\cite{Wign:70}, let us assume that some local hidden (LH) variable allows us to classify entangled pairs into various populations, 
according to the outcomes that $A$ and $B$ would find choosing to measure the spin along $\hat a$, $\hat b$ or $\hat c$. 
In any given event, the pair must be a member of one of the eight populations shown in table \ref{tab-Wig}. For instance, population 4 would correspond to $A$ obtaining $+1,-1,-1$ and $B$ obtaining $-1,+1,+1$ when measuring along one of the $\hat a$, $\hat b$ or $\hat c$ directions, respectively. Note how there is no population where $A$ and $B$ would obtain the same sign when measuring along the same direction, due to the conservation of angular momentum. Now, suppose that $A$ selects $\hat a$ and gets $+1$: if $B$ selects $\hat a$, the outcome must necessarily be $-1$, while if $B$ selects $\hat b$ or $\hat c$, the outcome can be $\pm 1$, giving rise to four distinct populations, $N_{1,2,3,4}$. The other four populations $N_{5,6,7,8}$ correspond to $A$ selecting $\hat a$ and getting $-1$.

\begin{table}[h!]
  \begin{center}
    \label{tab-pop}
    \begin{tabular}{c||c|c} 
      \textbf{Population} & \textbf{Particle 1 (A)} & \textbf{Particle 2 (B)} \\ \hline 
      $N_1$ & $(\hat a +, \hat b + , \hat c +)$ & $(\hat a - , \hat b  -, \hat c -)$ \\
      $N_2$ & $(\hat a +, \hat b + , \hat c -)$ & $(\hat a - , \hat b  -, \hat c +)$ \\
      $N_3$ & $(\hat a +, \hat b - , \hat c +)$ & $(\hat a - , \hat b  +, \hat c -)$ \\
      $N_4$ & $(\hat a +, \hat b - , \hat c -)$ & $(\hat a - , \hat b  +, \hat c +)$ \\
      $N_5$ & $(\hat a -, \hat b + , \hat c +)$ & $(\hat a + , \hat b  -, \hat c -)$ \\
      $N_6$ & $(\hat a -, \hat b + , \hat c -)$ & $(\hat a + , \hat b  -, \hat c +)$ \\
      $N_7$ & $(\hat a -, \hat b - , \hat c +)$ & $(\hat a + , \hat b  +, \hat c -)$ \\
      $N_8$ & $(\hat a -, \hat b - , \hat c -)$ & $(\hat a + , \hat b  +, \hat c +)$ \\
    \end{tabular}
    \caption{Particle populations according to Wigner's argument.}
    \label{tab-Wig}
  \end{center}
\end{table}

Suppose that $A$ and $B$ find $\alpha=\beta=+1$, measuring along $\hat a$ and $\hat b$ respectively. 
By inspecting table \ref{tab-Wig}, 
one sees that the particle pair must belong to either type 3 or type 4:
hence, the number of particle pairs for which this happens is $N_3+N_4$.
Because $N_i$ is positive semi-definite, we must have inequality relations such as:
\beq
N_3+N_4 \leq (N_2+N_4)+(N_3+N_7)\,. 
\label{eq-popW}
\eeq
Within this model the probability that, in a random selection,
$A$ and $B$ find $\alpha=\beta=+1$ measuring along $\hat a$ and $\hat b$ respectively, is given by:
\beq
P(\hat a +, \hat b +)_{LH}= \frac{N_3+N_4}{  \sum_{i=1}^{8}  N_i} \,\,  ,
\eeq
where $LH$ stand for local hidden.
Similarly, one has 
\beq
P(\hat a +, \hat c +)_{LH}= \frac{N_2+N_4}{\sum_i^8 N_i}\,\,\, ,\,\,\, P(\hat c +, \hat b +)_{LH}= \frac{N_3+N_7}{\sum_i^8 N_i} \,\,.
\eeq
Following Wigner's reasoning, on the basis of such an assumption, the previous inequality for populations, eq. (\ref{eq-popW}), becomes
\beq
P(\hat a +, \hat b +)_{LH} \le  P(\hat a +, \hat c +)_{LH} + P(\hat c +, \hat b +)_{LH} \,\,\, ,
\label{eq-BW}
\eeq
which is usually referred to as Bell-Wigner inequality.

Notice that Wigner's assumption of eight populations fits into a LH variable framework, because $A$’s result is predetermined independently of $B$’s choice of the measurement direction. 
A particular configuration of the experimental setup allows for a captivating visualization, as discussed in app. \ref{app-rhombo}, is the case in which $\hat a$, $\hat c$ and $\hat b$ are coplanar and form two angles of $\pi/4$.

We now introduce the quantity $Q^W$, defined as
\beq
Q^W = P(\hat a +, \hat b +) -  P(\hat a +, \hat c +) - P(\hat c +, \hat b +)  \,\,\, .
\label{def-QW} 
\eeq
For models based on LH variables, from eq. (\ref{eq-BW}) it follows that $Q^W_{LH} \le 0$. We now inspect whether $Q^W$ can be positive in models that do not rely on the assumption of LH variables.

Let us turn in particular to the standard interpretation of quantum mechanics: the first measurement which is carried out induces the collapse of the wave function of the entangled pair, thus violating Einstein's principle of locality. In any case, in the quantum mechanical framework, the above probabilities are explicitly calculable (see for instance \cite{Saku:85}), and turn out to be given by  
\beq
P(\hat a\pm, \hat b\pm)_{QM}=\frac{1}{2} \sin^2\left(\frac{\theta_{ab}}{2}\right) \,\,,
\,\, P(\hat a \pm, \hat b \mp)_{QM}=\frac{1}{2} \cos^2\left(\frac{\theta_{ab}}{2}\right) \,\, ,
\label{eq-prob}
\eeq
where $\theta_{ab}$ is the angle between $\hat a$ and $\hat b$.
For quantum mechanics, the quantity $Q^W$ of eq. (\ref{def-QW}) thus reads 
\beq
Q^W_{QM}= \frac{1}{2} \left( \sin^2\left(\frac{\theta_{ab}}{2}\right) -
 \sin^2\left(\frac{\theta_{ac}}{2}\right) - \sin^2\left(\frac{\theta_{cb}}{2}\right) \right)  \,\, .
 \label{eq-BW-QM}
\eeq
It is easy to show that, for certain configurations of angles, $Q^W_{QM}$ is positive, thus violating the Bell-Wigner inequality (\ref{eq-BW}). 
For instance, let us consider the case in which
$\hat a=\hat z$, $\hat b$ lies in the $xz$ plane, 
while $\hat c$ is a generic unit vector in spherical coordinates:
\beq
    \hat a=\hat z \,\,\, ,  \,\,\,
    \hat b = (\sin\theta_{ab}, 0,\cos\theta_{ab}) \,\,\, , \,\,\,
    \hat c = (\sin\theta_{ac} \cos\phi, \sin\theta_{ac} \sin\phi,\cos\theta_{ac})\,\,\,.
    \label{eq-vectors}
\eeq

\begin{figure}[h!]
\vskip .5cm 
 \begin{center}
    \includegraphics[width=7.5cm]{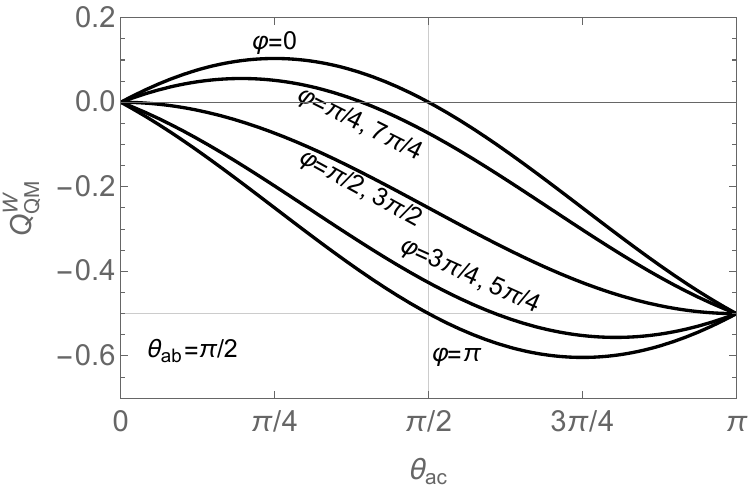}  \,\,\,
    \includegraphics[width=7.5cm]{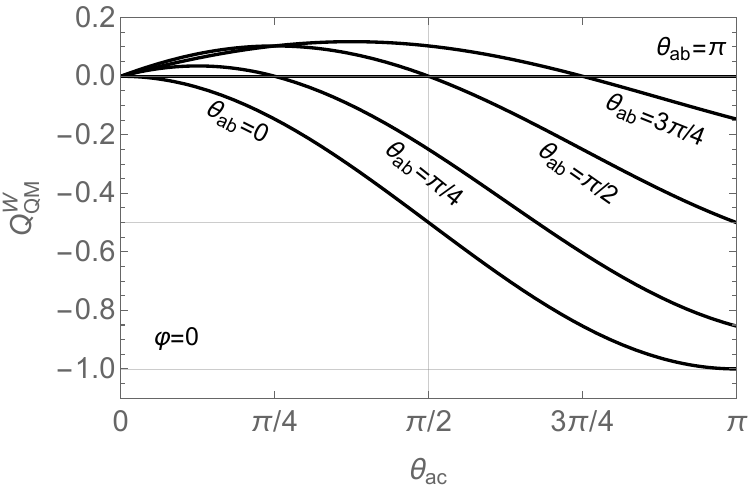}       
    \end{center}
\caption{\baselineskip=15 pt \small 
$Q^W_{QM}$ is shown as a function of $\theta_{ac}$. Left: for selected values of $\phi$ and taking $\theta_{ab}=\pi/2$. Right: for selected values of $\theta_{ab}$ and taking $\phi=0$.
}\label{fig-BW}
\vskip .5 cm
\end{figure}

First, consider the case $\theta_{ab}=\pi/2$ 
and let $\theta_{ac}$ vary in the range $[0,\pi]$, for selected values of $\phi$: the quantity $Q^W_{QM}$, is shown in the left panel of fig.~\ref{fig-BW}, where one can see that the violation of (\ref{eq-BW}) is maximal for $\phi=0$ and $\theta_{ac}=\pi/4$. Secondly, in the right panel of fig.~\ref{fig-BW}, we study $Q^W_{QM}$ focusing on the coplanar case $\phi=0$, for selected values of $\theta_{ab}$.

The maximal violation configuration ($\phi=0$, 
$\theta_{ac}=\pi/4$, $\theta_{ab}=\pi/2$) 
is the one that, as already mentioned, allows for a remarkable representation in terms of solids, as discussed in app. \ref{app-rhombo}.

The violation of the Bell-Wigner inequality (\ref{eq-BW}) due to quantum non-locality shows that interpreting entangled particles as they were classical objects carrying labels as in Wigner's argument, is too naive. Within the quantum mechanics approach, all states $\ket{\Psi^-}$ are equal (and indistinguishable); the violation of the above inequality tells that the spin components (along non-orthogonal directions) are not simultaneous elements of physical reality, and this questions local realism\,\footnote{$B$'s element of physical reality depends on the measurement direction chosen by $A$; as the measurement of $A$ actually affects the outcome of $B$, local realism is false. 
}. It has however to be mentioned that the debate on the actual implications of the violation of the Bell-Wigner inequality is still open \cite{Geu:1998, Scu:05, Hess:17, Hess:22}.

\subsection{Quantum circuits for the Bell-Wigner inequality}
\label{sub-qc-WB}

We turn now to a didactic tool to practically test the Bell-Wigner inequality.
We first discuss the related quantum circuit in an abstract way, and then show the code for the specific implementation in Qibo.

Let us give a very brief introduction to the basic concepts of quantum computing, useful for understanding what follows. The first difference between classical computers and a quantum computer relies on the way the information is stored in the device; classical computers codify any information into bit strings, namely, lists of bits assuming values 0 or 1. In contrast, a quantum computer is built using quantum bits: two-level quantum systems whose state $\ket{\psi}$ can assume any superposition of two basis vectors (called computational basis):
\begin{equation}
\ket{\psi} = \alpha_0 \ket{0} + \alpha_1 \ket{1}, \qquad \text{where} \qquad \alpha_0, \alpha_1 \in \mathbb{C} \qquad \text{and} \qquad |\alpha_0|^2 + |\alpha_1|^2 = 1.
\label{eq:qubit}
\end{equation}
Several practical implementations of qubits are currently explored, such as superconducting loops~\cite{superconducting}, trapped ions~\cite{trapped_ions}, neutral atoms~\cite{neutral_atoms} and photonics~\cite{photons}.

The notation of eq.~\eqref{eq:qubit} can easily be extended to the case of $n$ qubits, by defining the system state as a $2^n$-long vector of complex numbers fulfilling the proper normalization condition. The state of a system of qubits can be manipulated through unitary operators, defining a completely reversible computational setup.
One of the most common computational formulations of quantum computing is known as gate-based quantum computing, and consists in codifying any possible unitary operation as a sequence of quantum gates (which are, in practice, one- or two-qubit operations). A collection of gates applied to a system of qubits forms a quantum circuit, which is used to prepare a target final state given an initial state. Classical simulations of quantum computing operations typically assume the initial system state to be one in which all qubits are prepared in the computational zero state. Usually, this configuration corresponds to the ground state of the underlying physical system. This choice is common because preparing and maintaining this initial state is more practical when using an actual quantum computer: for example, superconducting qubits naturally occupy the physical state corresponding to the computational zero state when the chip temperature is sufficiently low, whereas an excited qubit maintained at such a temperature will spontaneously relax back into its ground state. Therefore, it is natural to consider the computational zero state as the initial state when describing the operation of a quantum computer.

For a more comprehensive introduction to quantum computing and for a better understanding of what follows, we suggest reading~\cite{nielsen00, mermin}.

Turning now to the Bell-Wigner inequality, in order to obtain the three probabilities entering eq. (\ref{def-QW}), one needs three quantum circuits, that is three different physical systems, as follows: a first circuit allows to compute $P(\hat a +, \hat b +)$, according to eq. (\ref{eq-vectors}), by evaluating $P(0,0)$:

\begin{equation}
\label{circ:had_test_diag}
\Qcircuit @C=1.2em @R=1em {
\lstick{|1\rangle}&\gate{H}&\ctrl{1}   \barrier[-.2em]{1} &\qw&\qw        & \meter & \cw & \rstick{A,\hat a} \\
\lstick{|1\rangle}& \qw     &\targ                                   &\qw&\gate{R_y(\theta_{ab})} \gategroup{2}{5}{2}{6}{.7em}{--} & \meter  & \cw & \rstick{B,\hat b} \\
 } \nonumber
\end{equation}
\vspace{0.5cm}

A second quantum circuit allows to measure $P(\hat a +, \hat c +)$ by evaluating $P(0,0)$:

\begin{equation}
\label{circ:had_test_diag}
\Qcircuit @C=1.2em @R=1em {
\lstick{|1\rangle}&\gate{H}&\ctrl{1}   \barrier[-.2em]{1} &\qw&\qw        & \meter & \cw & \rstick{A,\hat a} \\
\lstick{|1\rangle}& \qw     &\targ    &\qw &\gate{R_z(\phi)} \gategroup{2}{5}{2}{7}{.7em}{--} & \gate{R_y(\theta_{ac})} & \meter  & \cw & \rstick{B,\hat c} \\
 }  \nonumber
\end{equation}
\vspace{0.5cm}

\begin{minipage}{\textwidth}

Finally, a third circuit allows to measure $P(\hat c +, \hat b +)$ by evaluating once again $P(0,0)$:

\begin{equation}
\label{circ:had_test_diag}
\Qcircuit @C=1.2em @R=1em {
\lstick{|1\rangle}&\gate{H}&\ctrl{1} \barrier[-.2em]{1} &\qw &\gate{R_z(\phi)} \gategroup{1}{5}{1}{7}{.7em}{--} & \gate{R_y(\theta_{ac})} & \meter  & \cw & \rstick{A,\hat c} \\
\lstick{|1\rangle}& \qw     &\targ         &\qw &\gate{R_y(\theta_{ab})}                \gategroup{2}{5}{2}{6}{.7em}{--}. & \meter  & \cw & \rstick{B,\hat b} \\
 }  \nonumber
\end{equation}
\vspace{0.5cm}
\end{minipage}

All circuits start by producing the same (at least in principle) Bell state, the $s=0$ singlet of eq.\,(\ref{eq-singlet}); 
using the standard Bloch sphere's notation for the qubits, the latter reads: $|\Psi^{-}\rangle=(|01\rangle-|10\rangle)/\sqrt{2}$.
Spin eigenstates along $\hat z$, $|+\rangle$ and $|-\rangle$, indeed correspond to $|0\rangle$ and $|1\rangle$ respectively.
The required Bell state is obtained at the position indicated by the vertical dashed line, starting with two $|1\rangle$ qubits, applying an Hadamard ($H$) gate on the first qubit $q_0$, and then a CNOT gate to the second qubit $q_1$ controlled by $q_0$.

As for the measurement directions, we reproduce the setup of (\ref{eq-vectors}). A measurement along $\hat z$ is simply the standard measurement operation. For a measurement along $\hat c$, one has to first carry out a rotation by $\phi$ along $\hat z$, followed by a rotation by $\theta_{ac}$ along $\hat y$, which are implemented by the $R_z(\phi)$ and $R_y(\theta_{ac})$ gates (this block is represented by the dashed rectangle). For $\hat b$, a rotation by $\theta_{ab}$ along $\hat y$ is enough. The outcomes $(0,0)$ correspond to $\alpha=\beta=+1$; and for the three circuits above, their frequencies are respectively equal to the probability values $P(\hat a +, \hat b +)$, $P(\hat a +, \hat c +)$ and $P(\hat c +, \hat b +)$ entering eq. (\ref{def-QW}).

From a didactic perspective, exploring such configuration helps discussing two issues. Firstly, from fig.~\ref{fig-BW} one expects maximal violation to take place when the three vectors are coplanar, whereas the violation reduces by increasing $\phi$. Secondly, it is well known that the order in which the quantum circuit performs rotations is important: since the measurement operation is conventionally always along $\hat z$, performing a $R_z$ rotation right before the measurement would produce no measurable effect.

\subsection{Implementation of the Bell-Wigner inequality in Qibo}
\label{sub-qibo-WB}

Qibo~\cite{qibo_paper} is an open-source software framework for quantum computing offering a simple Python interface to execute quantum circuits, both as a classical simulation and on real quantum computers~\cite{Efthymiou_2024, pasquale2024qibocalopensourceframeworkcalibration}. Qibo provides different runtime backends~\cite{Efthymiou_2022, pasquale2024statevectorsimulationqibo} that make use of CPUs as well as GPUs when available, such as to obtain significant speed-ups when simulating larger circuits ({\it e.g.} with more than 20 qubits).
Various approaches exist in order to simulate the evolution of a quantum system composed of $n$ qubits. Appendix \ref{app-qibo} describes how this is implemented in Qibo.

Focusing on the Bell-Wigner quantum circuits, we note that Qibo follows the standard assumption where all qubits' initial states are set to be $|0\rangle$ by default, therefore the initial $|1\rangle$ states can be prepared by acting on $|0\rangle$ with an $X$ gate. A suitable circuit to simulate the Bell-Wigner inequality can then be written as follows:

\vspace{0.4cm}
\begin{lstlisting}[language=Python]
from qibo import Circuit, gates
from qibo import set_backend

# Set the most suitable backend for the execution
set_backend(backend="numpy")

# Instantiate a generic circuit for Bell inequalities
c = Circuit(2)
c.add(gates.X(q=0))
c.add(gates.X(q=1))
c.add(gates.H(0))
c.add(gates.CNOT(q0=0, q1=1))

# Add parameterized rotations and measurements
c.add(gates.RZ(q=0, theta=phi_0))
c.add(gates.RZ(q=1, theta=phi_1))
c.add(gates.RY(q=0, theta=th_0))
c.add(gates.RY(q=1, theta=th_1))
c.add(gates.M(0, 1))
\end{lstlisting}
\vspace{0.4cm}

This code implements the following circuit:

\begin{equation}
\label{circ:had_test_diag}
\Qcircuit @C=1.2em @R=1em {
\lstick{|0\rangle}&\gate{X}&\gate{H}&\ctrl{1} \barrier[-.2em]{1} & \qw & \gate{R_z(\phi_0)} & \gate{R_y(\theta_0)} & \meter & \cw & \\
\lstick{|0\rangle}&\gate{X}& \qw & \targ & \qw & \gate{R_z(\phi_1)} & \gate{R_y(\theta_1)} & \meter & \cw \\
} \nonumber
\end{equation}
\vspace{0.5cm}

For the simulation, we first explore the coplanar case by setting $\phi_0 = \phi_1 = 0$, whereas relevant values of $\theta_0$ and $\theta_1$ are: $(\theta_0, \theta_1) = (0, \theta_{ab})$ to evaluate $P(\hat a+, \hat b+)$; $(\theta_0, \theta_1) = (0, \theta_{ac})$ to evaluate $P(\hat a+, \hat c+)$; and finally ($\theta_0, \theta_1) = (\theta_{ac}, \theta_{ab})$ to evaluate $P(\hat c+, \hat b+)$. In particular, evaluating any such $P$ corresponds to counting the cases where both qubits are measured as \texttt{0}, which is represented in the code by:

\vspace{0.4cm}
\begin{lstlisting}[language=Python]
# Frequentist approach to measure P
P = c(nshots=nshots).frequencies()["00"] / nshots
\end{lstlisting}
\vspace{0.4cm}

Afterwards, the simulation explores the case of a rotation of $\hat c$ along the $\hat z$ axis, with a $\phi > 0$ angle with respect to the $\hat a,\hat b$ plane: to account for such rotation, we fix $\theta_{ab} = \pi/2$, and we set $(\phi_0, \phi_1) = (0, \phi)$ to evaluate $P(\hat a+, \hat c+)$; and $(\phi_0, \phi_1) = (\phi, 0)$ to evaluate $P(\hat c+, \hat b+)$.

\begin{figure}[t!]
    \centering
    \includegraphics[width=0.72\linewidth]{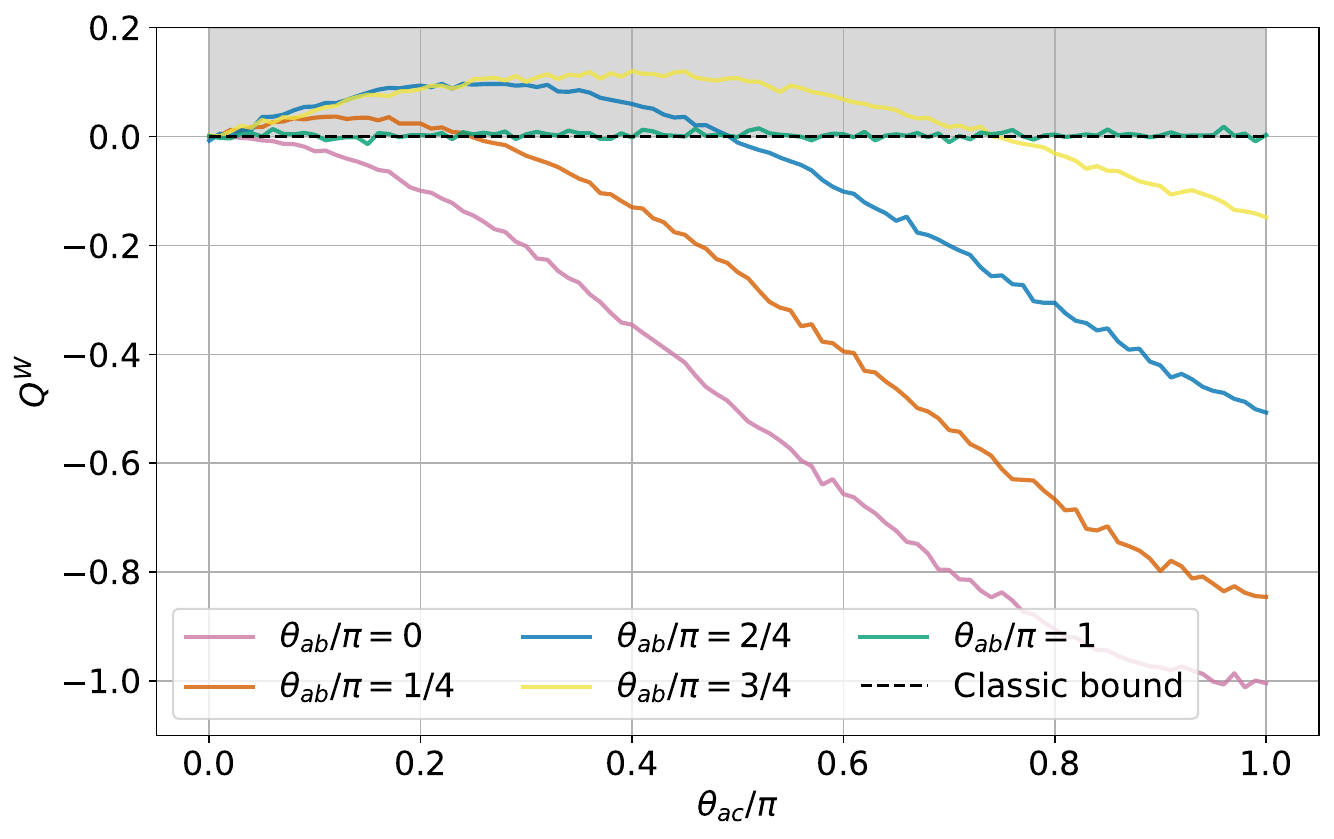}
    \vskip 0.5cm
    \includegraphics[width=0.72\linewidth]{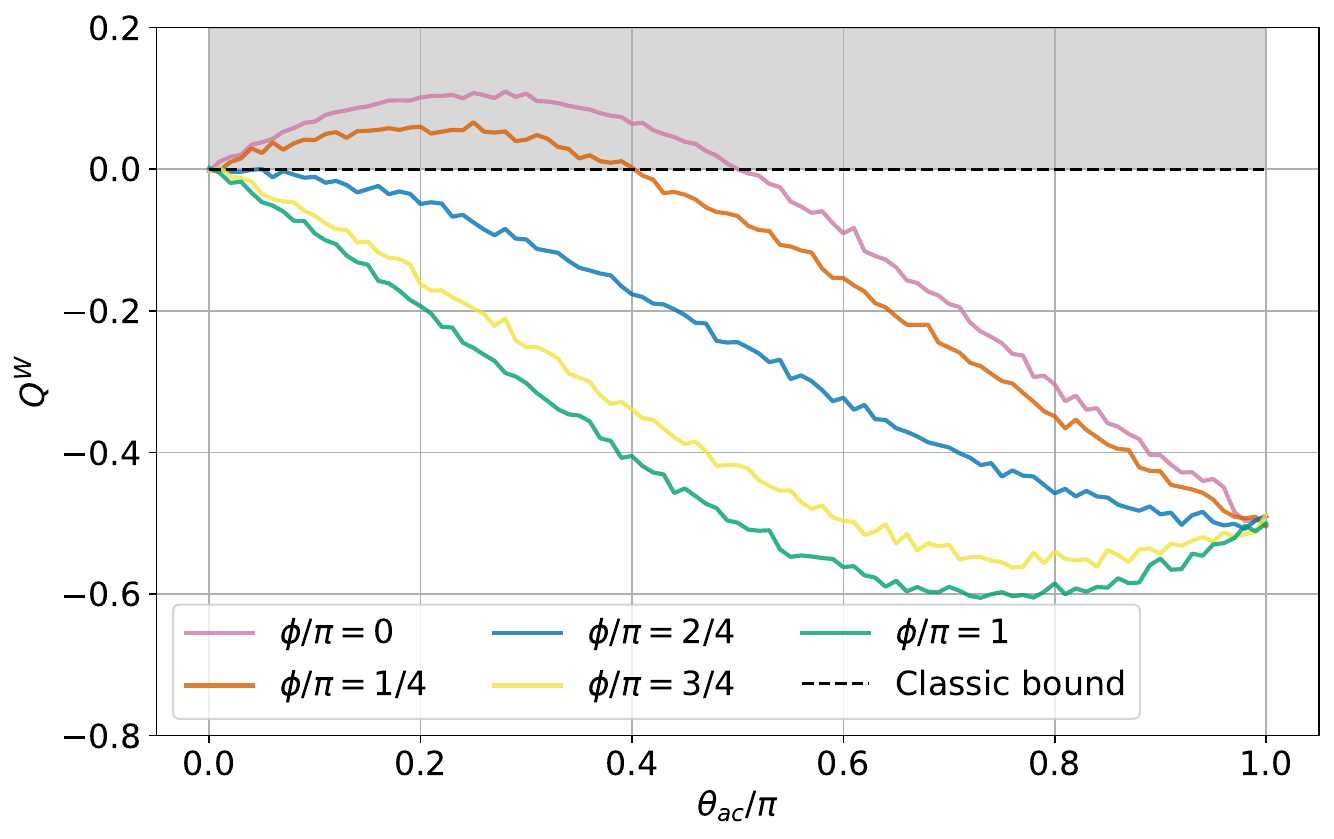}
    \caption{Qibo simulation for the Bell-Wigner inequality (\ref{eq-BW}). The $Q^W$ value is shown as a function of $\theta_{ac}$, and the gray band corresponds to the violation region. Top: for selected values of $\theta_{ab}$ and taking $\phi=0$. Bottom: for selected values of $\phi$ and taking $\theta_{ab}=\pi/2$.}
    \label{fig:qibo-bw}
\end{figure}

The plots in fig.~\ref{fig:qibo-bw} show the value of $Q^W$, as defined by eq. (\ref{def-QW}) and as computed by our simulation. We execute $N_{\rm shots} = 10,000$ runs for each chosen set of parameters: the running variable is always $\theta_{ac}$, as $\hat c$ represents our probe in the experimental setting. The multiple lines are computed on one hand setting different values for $\theta_{ab}$ with a fixed value for $\phi$, and on the other hand setting different values for $\phi$ with a fixed value for $\theta_{ab}$. All angles run in the range $[0, \pi]$.

The $(\theta_{ab}, \theta_{ac})$ values (resp. the $(\phi, \theta_{ac})$ values) for which the $Q^W$ quantity is strictly positive represent all the cases where the Bell-Wigner inequality is violated by the quantum system, and are emphasized in grey. Qibo results of course coincide with quantum mechanics predictions of fig.\,\ref{fig-BW} discussed previously. The simulation confirmed the theoretical findings whereby the Bell-Wigner inequality is violated, when $\phi = 0$, for $\theta_{ac} < \theta_{ab}$, and it is conserved for $\theta_{ac} \ge \theta_{ab}$, for any $\theta_{ab}$ chosen in the range $]0, \pi[$.

\section{The original Bell inequality} 
\label{sec-B}

The original Bell inequality \cite{Bell:64} is not based on probabilities as the Bell-Wigner version, but rather on correlations of the results by observers $A$ and $B$. The setup is however exactly the same as for Bell-Wigner.

\subsection{Theoretical framework}
\label{sub-thfr-B}

In the previous discussion, we denoted the possible results of $A$ measuring the spin along $\hat a$ by $\alpha$, which can thus be equal to $\pm 1$; similarly, $B's$ results are $\beta=\pm 1$.
Let us consider the quantity $\alpha \beta$, that is the product of the results for a single pair. 
The correlation is defined as the mean value of such product over many pair measurements, $C(\hat a, \hat b) = \overline{\alpha \beta}$. Hence, it can be calculated in terms of the probabilities defined in the previous section:
\beq
C(\hat a, \hat b) 
= P(\hat a +, \hat b +) -P(\hat a +, \hat b -)-P(\hat a -, \hat b +)+P(\hat a -, \hat b -)\,\, .
\label{eq-Corr}
\eeq

Bell \cite{Bell:64} showed that a local hidden variable model would satisfy
\beq
| C(\hat a, \hat b)_{LH}- C(\hat a, \hat c)_{LH} | \leq 1+ C(\hat c, \hat b)_{LH}\,\, ,
\label{eq-B64}
\eeq
whose derivation is reviewed in app. \ref{app-Bell}.

We now define the quantity $Q^B$, related to Bell inequality, as
\beq
Q^B = | C(\hat a, \hat b)- C(\hat a, \hat c)  | - C(\hat c, \hat b)  
\label{def-QB}
\eeq
and inspect if it can be larger than one for models not relying on LH variables.

Within quantum mechanics, the above correlation for a singlet state is easily calculable (see {\it e.g.} \cite{Fano:17}).
Substituting eq. (\ref{eq-prob}) in eq. (\ref{eq-Corr}), one has 
\beq
C(\hat a, \hat b)_{QM}=-\hat a \cdot \hat b =-\cos\theta_{ab}\,\,. 
\label{eq-C-QM}
\eeq
The quantity of eq. (\ref{def-QB}) thus becomes
\beq
Q^B_{QM}=|- \cos\theta_{ab} + \cos\theta_{ac}| +\cos\theta_{cb} \,\, .
\label{eq-B64-QM}
\eeq
For instance, taking $\hat a=\hat z$, $\hat b=\hat x$ and $\theta_{ac}=\pi/4$, the inequality is violated, as $Q^B_{QM}=\sqrt{2}$, which is indeed larger than one.

\begin{figure}[h!]
\vskip .5cm 
 \begin{center}
    \includegraphics[width=7.6cm]{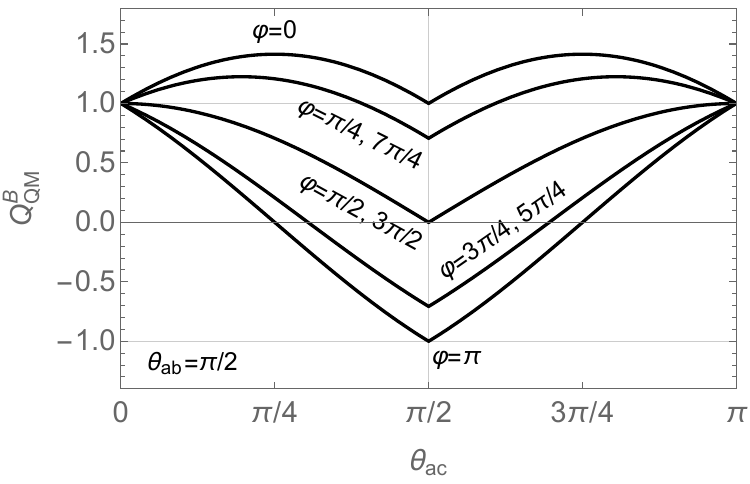} \,\, \,\,\,\includegraphics[width=7.6cm]{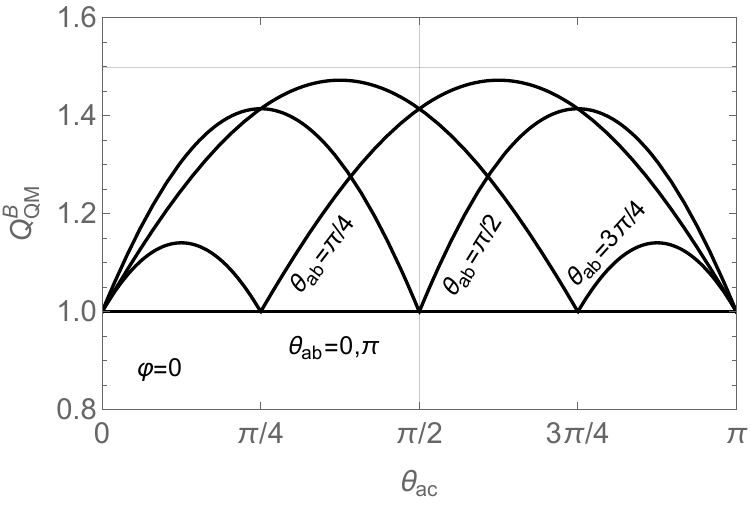}       
 \end{center}
\caption{\baselineskip=15 pt \small 
$Q^B_{QM}$ is shown as a function of $\theta_{ac}=\theta$. Left: for selected values of $\phi$ and taking $\theta_{ab}=\pi/2$. Right: for selected values of $\theta_{ab}$ and taking $\phi=0$.
}\label{fig-B64}
\vskip .5 cm
\end{figure}

The left panel of fig.~\ref{fig-B64} shows $Q^B_{QM}$ in the case $\theta_{ab}=\pi/2$, for fixed values of $\phi$ and varying $\theta_{ac}$;
for $\phi=0$, the violation always occurs, apart from the trivial cases in which two unit vectors coincide. 
The right panel shows $Q^{B}_{QM}$ for selected values of $\theta_{ab}$, keeping $\phi=0$.

It has to be stressed that the Bell inequality assumes perfect anti-correlation, as discussed in app. \ref{app-Bell}: if one particle is measured along a given axis with spin-up, the other must be spin-down. This is crucial because local hidden variable theories predict deterministic outcomes under identical settings. However, real experiments like the ones executed on a quantum device, face imperfections—state preparation, measurement inefficiencies, and noise—making perfect anti-correlation unachievable. The anti-correlation requirement is avoided in the subsequent CHSH versions of Bell's inequality, which rely just on the factorizability condition for probabilities, more robust for experimental tests of quantum non-locality. For a detailed discussion on this subject, see {\it e.g.} \cite{scholar}.

\subsection{Implementation of the Bell inequality in Qibo}
\label{sub-qibo-B}

To simulate the Bell inequality we use the same circuit as in the previous module. However, we need to compute correlations of the form $\overline{\alpha\beta}$, which require the frequencies of all outcomes of the two qubits \cite{simcode}:

\begin{lstlisting}[language=Python]
    # Compute correlations
    freqs = c(nshots=nshots).frequencies()
    C = (freqs['00'] - freqs['01'] - freqs['10'] + freqs['11']) / nshots
\end{lstlisting}

\begin{figure}[h!]
    \centering
    \includegraphics[width=0.72\linewidth]{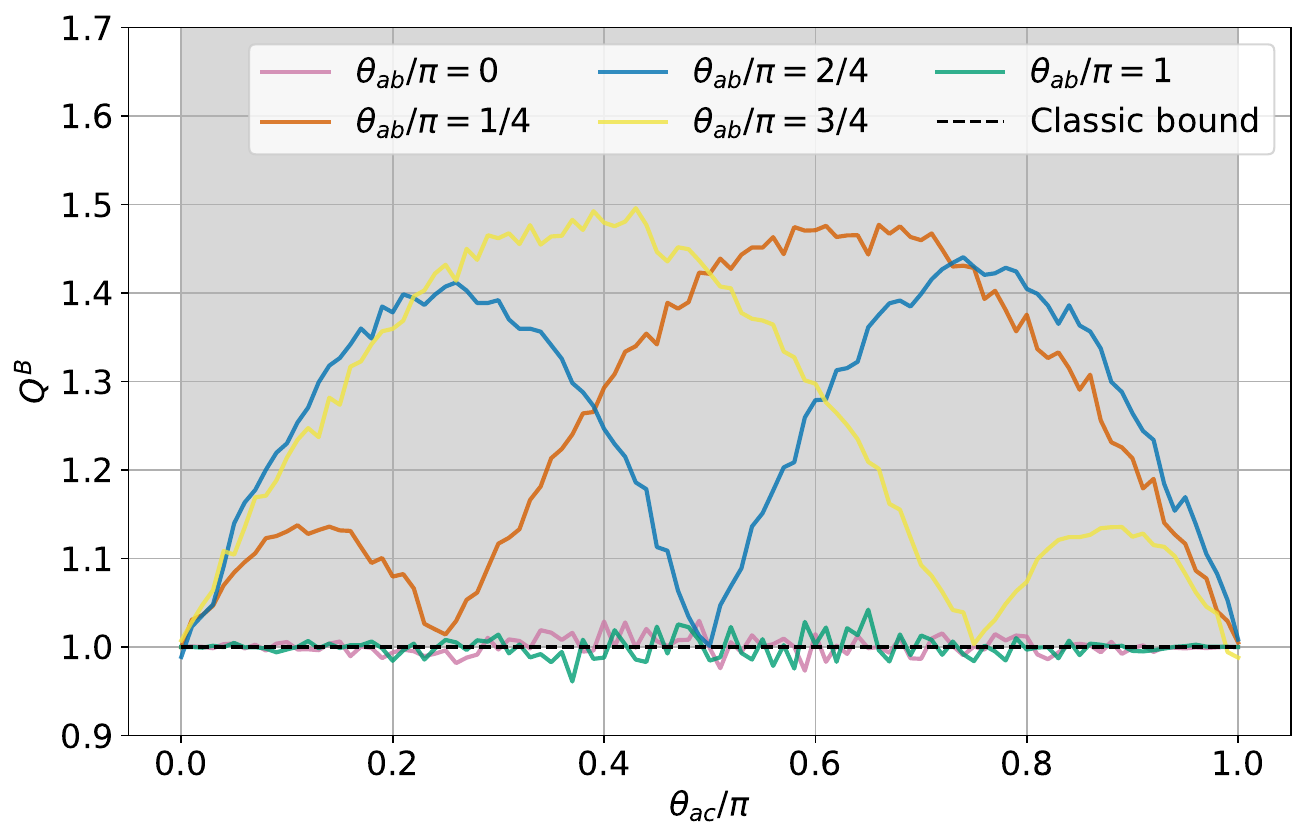}
    \vskip 0.5cm
    \includegraphics[width=0.74\linewidth]{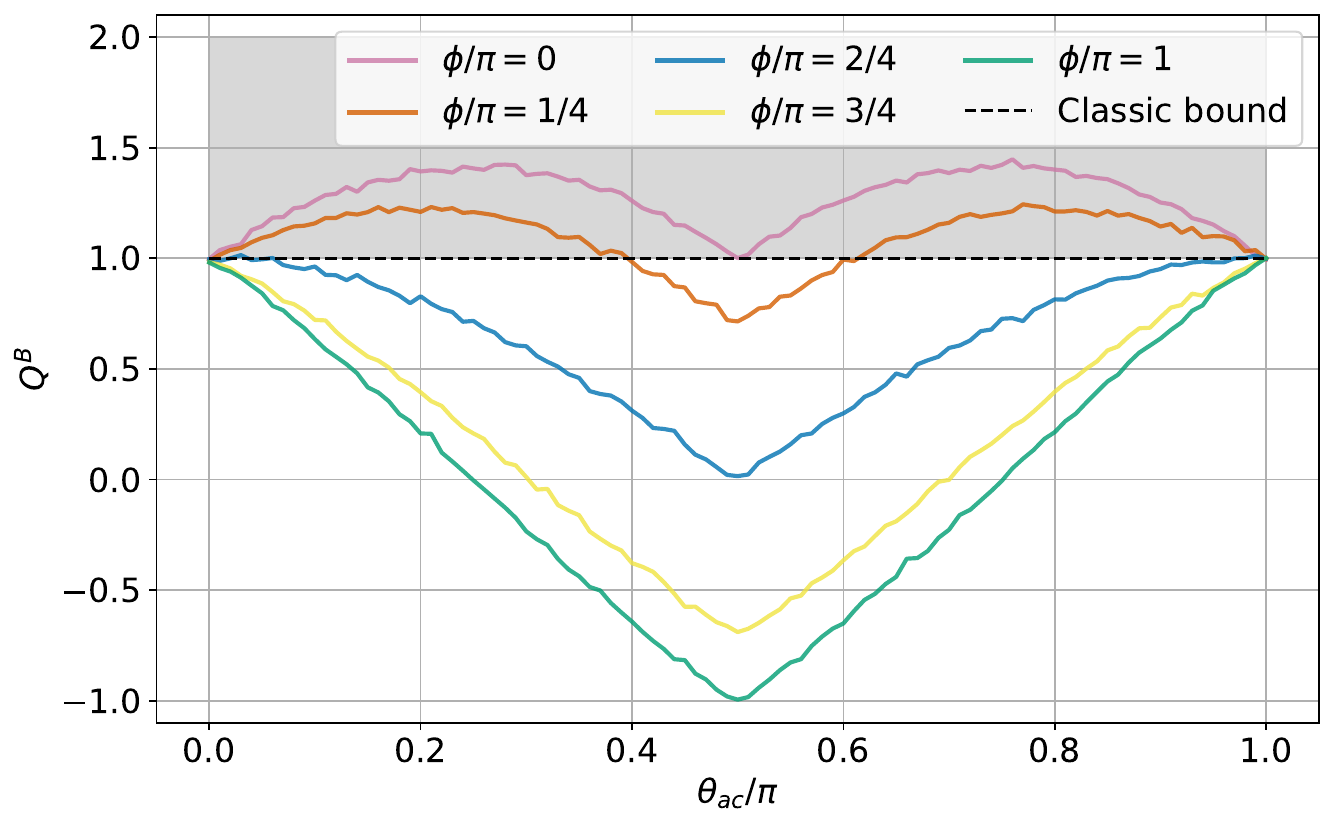}
    \caption{Qibo simulation for the Bell 1964 inequality (\ref{eq-B64}). The $Q^B$ value is shown as a function of $\theta_{ac}$, and the gray band corresponds to the violation region. Top: for selected values of $\theta_{ab}$ and taking $\phi=0$. Bottom: for selected values of $\phi$ and taking $\theta_{ab}=\pi/2$. }
    \label{fig:qibo-bell}
\end{figure}

The quantity of eq. (\ref{def-QB}) is thus
\beq
Q^B=| \overline{ \alpha \beta} - \overline{\alpha \gamma}   |- \overline{ \beta \gamma}    
\,\, ,
\eeq
whose results are shown in fig.\,\ref{fig:qibo-bell} as a function of $\theta_{ac}$ and for selected values of $\theta_{ab}$ and $\phi$ respectively; the regions where it exceeds $1$ are those where a violation occurs. 
In accordance to the theoretical results of fig.\,\ref{fig-B64}, if $\phi = 0$ a violation happens everywhere, apart trivial cases where two directions coincide, whereas as $\phi$ increases, the inequality is less and less violated.

\section{The CHSH inequality} 
\label{sec-CHSH}

As already mentioned, a setup that avoids the inconvenience of the anti-correlation assumption as in Bell's original inequality is the one denoted CHSH, for Clauser-Horne-Shimony-Holt \cite{CHSH:69}. Educational material related to this subject is already available \cite{Qiskit:CHSH, Qibo:CHSH}.
Here we explore with more details the results obtained for different possible spatial configurations of the experimental setup.

\subsection{Theoretical framework}
\label{sub-thfr-CHSH}

The setup is now such that observer $A$ performs measurements along directions $\hat a$ and $\hat b$, obtaining outcomes $\alpha$ and $\beta$ respectively; while $B$ along $\hat c$ and $\hat d$, obtaining outcomes $\gamma$ and $\delta$ respectively.  
An interesting combination of product of results is \cite{CHSH:69, Fano:17}:
\beq
S= \alpha (\gamma - \delta) + \beta (\gamma +\delta) \,\, .
\eeq

Since either $(\gamma - \delta)$ or $(\gamma + \delta)$ vanishes, while the other is equal to $\pm 2$, 
turning to correlations within a LH variable model, one obtains an inequality of the CHSH-type \cite{CHSH:69,Fano:17}
\beq 
|\overline{S}|_{LH}=
|C(\hat a, \hat c)_{LH}-C(\hat a, \hat d)_{LH}+C(\hat b, \hat c)_{LH}+C(\hat b, \hat d)_{LH}| \leq 2 \,\, .
\label{eq-CHSH}
\eeq

We thus define the related quantity 
\beq 
Q^S = |\overline{S}|=
|C(\hat a, \hat c)-C(\hat a, \hat d)+C(\hat b, \hat c)+C(\hat b, \hat d)|  \,\, ,
\label{def-QS}
\eeq
and inspect if it can be larger than $2$ in a model not relying on LH variables.

According to quantum mechanics, exploiting eq. (\ref{eq-C-QM}), the quantity above would read 
\beq
Q^{S}_{QM}=|- \cos\theta_{ac} + \cos\theta_{ad} - \cos\theta_{bc} -\cos\theta_{bd}| \,\, .
\label{eq-CHSH-QM}
\eeq

\begin{figure}[h!]
\vskip .5cm 
 \begin{center}
    \includegraphics[width=7.5cm]{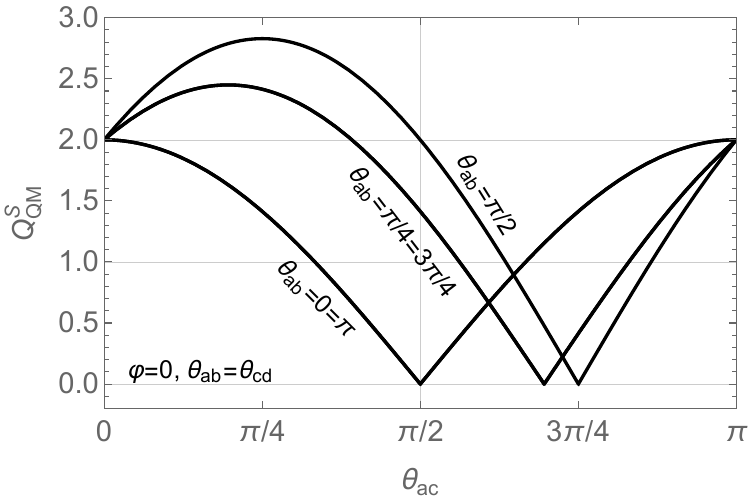}  \,\,\,  
    \includegraphics[width=7.5cm]{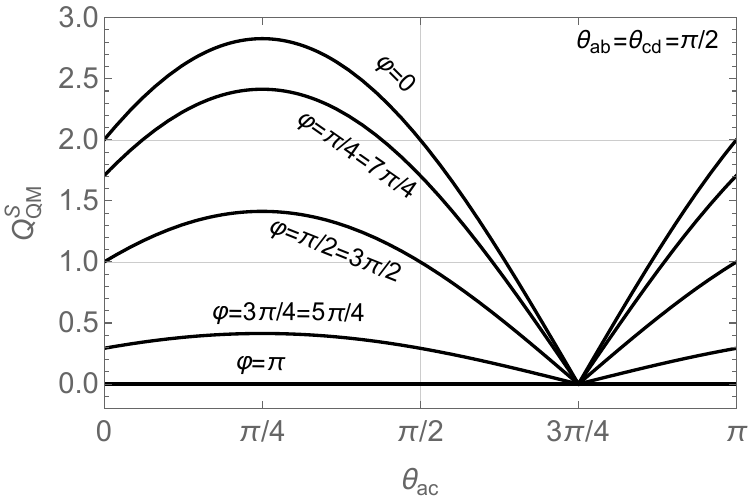} 
    \end{center}
\caption{\baselineskip=15 pt \small 
$Q^S_{QM}$, that is the left-hand side of CHSH inequality (\ref{eq-CHSH-QM}) according to quantum mechanics, is shown as a function of $\theta_{ac}$. Left: taking $\phi=0$ and selected values of $\theta_{ab}=\theta_{cd}$. Right:
taking $\theta_{ab}=\theta_{cd}=\pi/2$ and selected values of $\phi$.
}\label{fig-CHSH}
\vskip .5 cm
\end{figure}

The configuration which maximally violates the CHSH inequality 
(\ref{eq-CHSH}) is the one in which all measurement directions are coplanar and angles of $\pi/4$ are formed between $A$ and $B$'s directions, so that $\hat a=\hat z$, $\hat b=\hat x$, while $\hat c$ and $\hat d$ are orthogonal, with $\theta_{ac}=\pi/4$ and $\theta_{ad}=3\pi/4$; indeed, in this case the inequality would read $Q^{S}_{QM}=2 \sqrt{2} \leq 2$. The dependence on $\theta_{ac}$ is shown in fig.~\ref{fig-CHSH} for various configurations of the setup.

\subsection{Implementation of the CHSH inequality in Qibo}
\label{sub-qibo-CHSH}

For this case, we reuse the same circuit and correlation expression previously introduced for the Bell case, and compute the $Q^S$ quantity instead.

\begin{figure}[h!]
    \centering
    \includegraphics[width=0.72\linewidth]{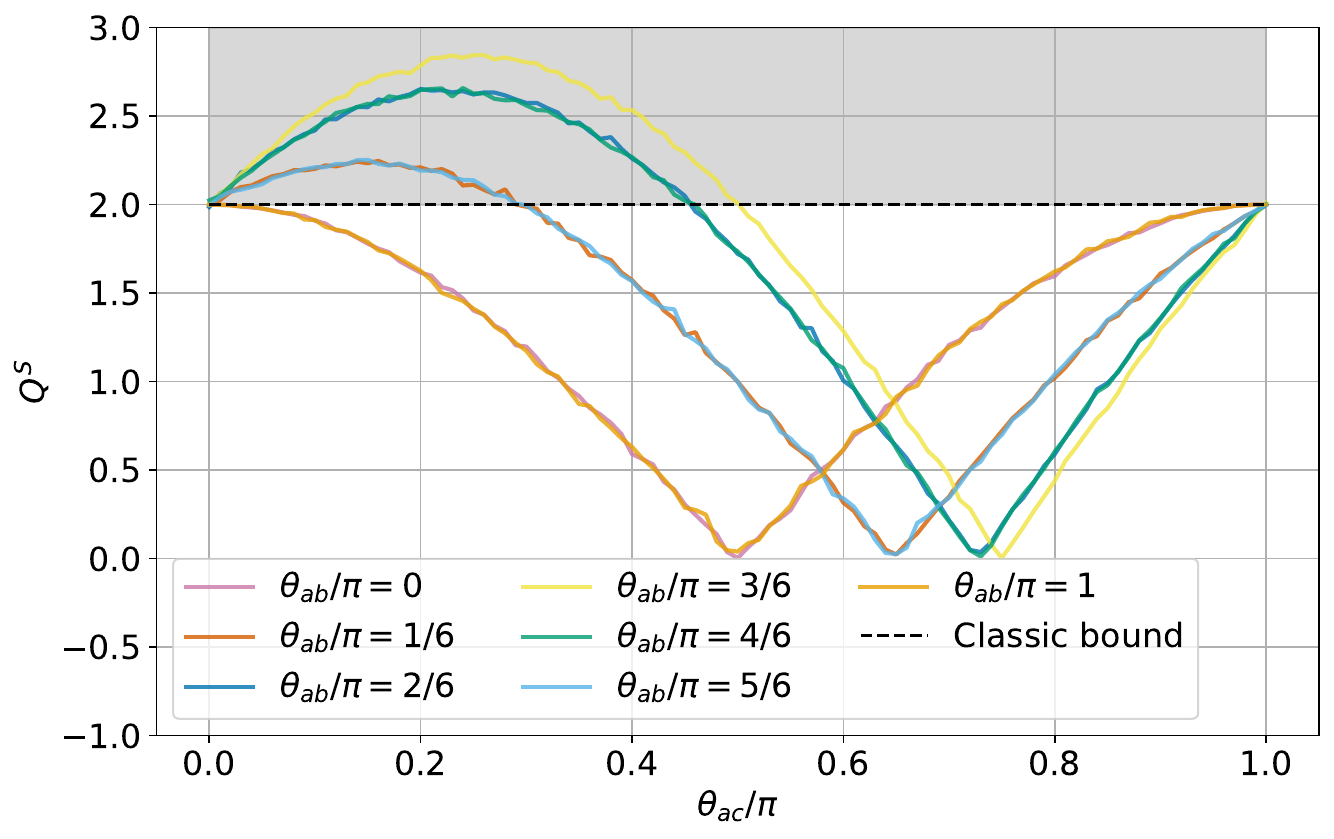}
    \vskip 0.5cm
    \includegraphics[width=0.72\linewidth]{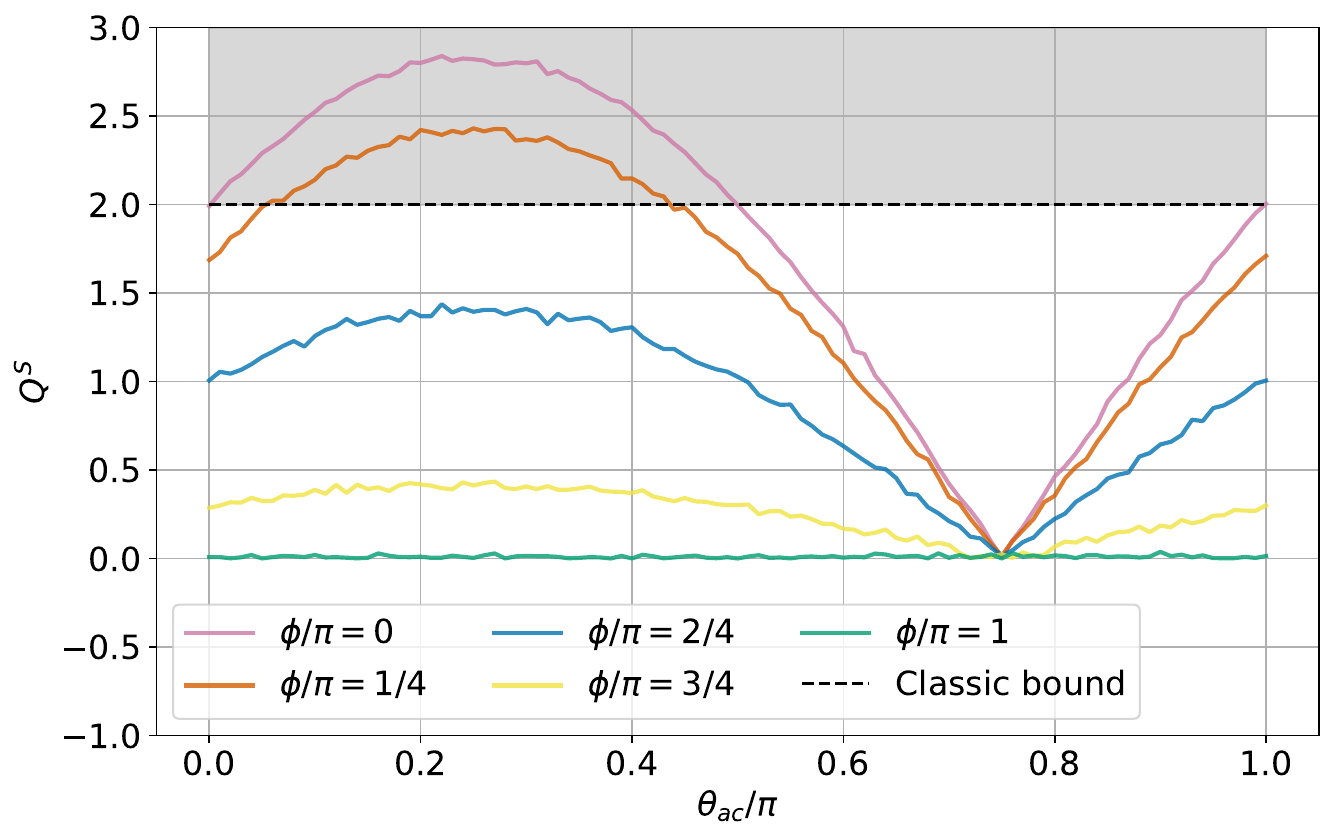}
    \caption{Qibo simulation of the CHSH inequality. The $Q^S$ value is shown as a function of $\theta_{ac}$, and the grey band corresponds to the violation region.
    Top: for selected values of $\theta_{ab} = \theta_{cd}$, with $\phi = 0$. Bottom: for selected values of $\phi$, taking $\theta_{ab} = \theta_{cd} = \pi/2$.}
    \label{fig:qibo-chsh}
\end{figure}

The results of this simulation are shown in fig.~\ref{fig:qibo-chsh}, for different values of the $\theta_{ab}$ parameter. As expected, the inequality is maximally violated for $\theta_{ab} = \pi/2$ and $\theta_{ac} = \pi/4$. A similar result is obtained when introducing the rotation of a $\phi > 0$ angle about the $z$ axis for $\hat c$ and $\hat d$ with respect to the $\hat a, \hat b$ plane: the inequality is less and less violated as $\phi$ grows from $0$ to $\pi$.

\begin{figure}
    \centering
    \includegraphics[width=0.55\linewidth]{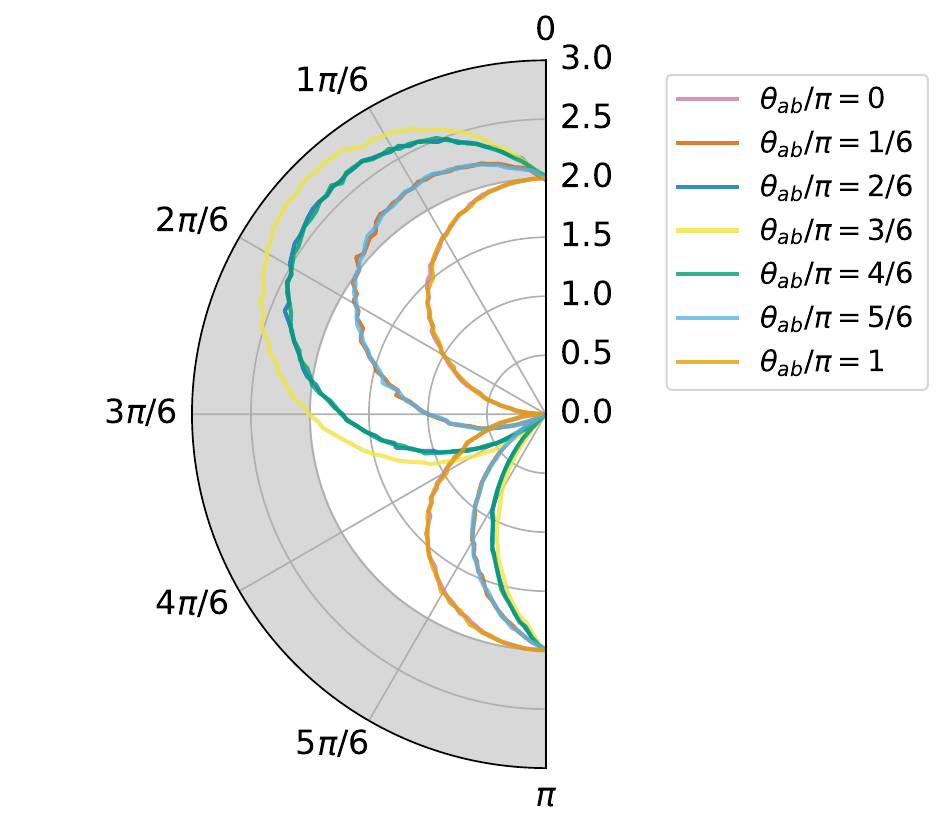}
    \caption{Polar plot of the Qibo simulation for the CHSH inequality, for selected values of $\theta_{ab} = \theta_{ac}$. Similarly to previous figures, the gray band corresponds to the violation region.}
    \label{fig:qibo-chsh-polar}
\end{figure}

Additionally, for this most popular inequality we propose a polar representation in fig.~\ref{fig:qibo-chsh-polar}, in order to better illustrate the running of the free variable $\theta_{ac}$: indeed, this value runs from 0, which corresponds to the $\hat c$ vector aligned to $\hat z$, conventionally represented as the North direction, to $\pi$, which corresponds to $\hat c$ aligned to the South direction. Similar polar plots have been included in all Python notebooks \cite{simcode}.

\section{Discussing Statistics and Noise}
\label{sec-sn}

In quantum computing experiments, the results are influenced by two primary sources of error: statistical fluctuations and intrinsic noise from the hardware. The statistical error arises due to the finite number of measurement shots ($N_{\rm shots}$), which affects the smoothness of the probability distribution and introduces "ripples" in the measured outcomes. Estimating the statistical error can be achieved through standard error analysis, where the uncertainty in the measurement is inversely proportional to the square root of the number of shots. As the number of shots increases, the statistical error decreases, making it possible to estimate the threshold at which this error becomes comparable to the device noise. In practice, determining the optimal number of shots requires balancing the trade-off between minimizing statistical error and mitigating noise, with diminishing returns observed as the number of shots continues to increase.

\subsection{Statistical Error}
 In quantum computing, when measuring a quantum state, we repeat the experiment multiple times (denoted as $N_{\rm shots}$) to obtain statistical confidence in the results.
 Let’s assume we are measuring the probability $P$ of a particular outcome ({\it e.g.}, finding the qubit in the $\ket{1}$ state). If we measure this over $N_{\rm shots}$ trials, the number of times this outcome is observed is $k$. The frequentist probability $\hat{P}$ of measuring the state above is given by:

 \begin{equation}
      \hat{P} = \frac{k}{N_{\rm shots}} \,\,.
 \end{equation}
 
The statistical error (standard deviation) of this binomial distribution can be approximated using standard error propagation for probabilities, which is:
\begin{equation}
    \sigma_{\rm stat} = \sqrt{\frac{P(1-P)}{N_{\rm shots}}}\,\,,
\end{equation}
that for large $N_{\rm shots}$ decreases as $\sigma_{\rm stat} \approx 1/ \sqrt{N_{\rm shots}}$.
Thus, by increasing the statistics, we reduce the statistical error, but the benefit diminishes as the rate of improvement scales with this factor.

\subsection{Noise Error}
The noise error, on the other hand, is intrinsic to the hardware and depends on various factors such as gate fidelity, decoherence, and readout errors. Let's denote this noise error as $\sigma_{\rm noise}$, which is typically characteristic of the quantum device. This value is independent of $ N_{\rm shots}$, meaning increasing it does not reduce the hardware noise.

In conclusion, the total error in the probability measurement can be estimated by combining the statistical error and noise error. If the two errors are independent, they can be added in quadrature:
\begin{equation}
    \sigma_{\rm total} = \sqrt{\sigma_{\rm stat}^2 + \sigma_{\rm noise}^2} \,\,.
\end{equation}
In a first approximation we cannot control the noise error, so this forces us to ensure that the statistical error is comparable to the noise error, so that we can estimate the required number of shots:
\begin{equation}
    N_{\rm shots} \approx \frac{P(1-P)}{\sigma^2_{\rm noise}}\,\,.
\end{equation}
This equation tells us how many shots are needed to make the statistical error comparable to the noise error. If the noise is significant, a smaller number of shots is required to achieve a similar level of statistical precision.
 
\subsection{Noise Sources in Quantum Computing}
Simulations of quantum circuits often serve as an idealized reference point, free from hardware-specific noise and imperfections. When comparing simulation results to those obtained from real quantum devices, discrepancies arise primarily due to decoherence, gate errors, and readout noise present in physical systems. 

\begin{figure}[h!]
    \centering
\includegraphics[width=1\linewidth]{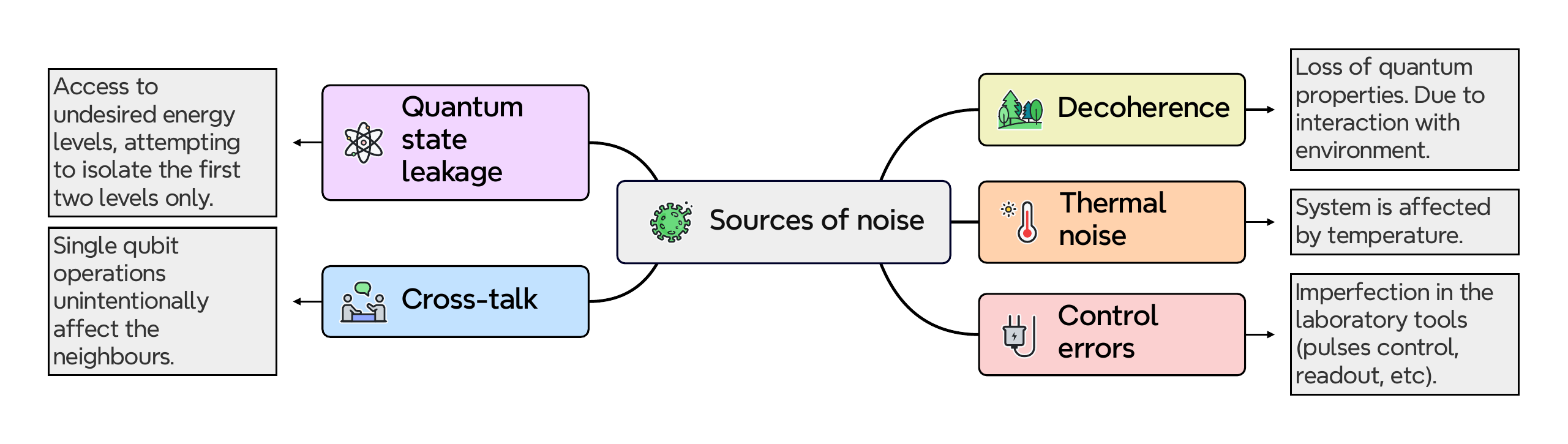}
    \caption{Some of the possible sources of noise of a quantum device.}
    \label{fig:noises}
\end{figure}

Simulations assume perfect quantum operations, providing a baseline for expected outcomes under ideal conditions. By contrasting these results with data from real devices, it becomes possible to quantify the impact of hardware noise and assess the device’s fidelity. While simulations can model noise to some extent using error channels, real-world behaviour is often more complex, and thus, testing on physical devices remains crucial for understanding practical limitations and improving error mitigation strategies.

As highlighted in fig.~\ref{fig:noises}, quantum computers are highly susceptible to various types of noise due to their sensitivity to environmental disturbances and imperfections in hardware components. These noise sources degrade the fidelity of quantum operations, causing errors in computations. The most common sources of noise include: decoherence, gate errors, readout errors, cross talks and all the errors related to the preparation and calibration of the qubits.
All these noise sources contribute to the total noise error $\sigma^2_{\rm noise}$ discussed earlier. They affect the quantum state, gate operations, and final measurements, thereby setting a practical limit on the accuracy of quantum computations. Since these errors are hardware-specific, their mitigation requires a combination of better hardware design, error-correction codes, and optimal calibration of quantum devices.

\section{Conclusion and overview}
\label{sec-conc}

Qubits represent a practical tool to get hands-on in quantum mechanics, in order to facilitate the understanding of fundamental issues of quantum systems and their intrinsic nature, which is yet subject of active debate in the scientific community.

Here we presented a thorough review of a number of well-known Bell-type inequalities, from the conceptually easier Wigner version, to the original inequality by Bell and its successive CHSH version.
A module is devoted to each one of those inequality. After a short theoretical presentation, the quantum circuit and the associated Qibo code are discussed~\cite{simcode}. These tools allow for an exploration of the possible configurations of the experimental setup, aimed at visualizing the transition between the classical and the quantum regimes. Here we discussed some original plots as an example, and the software is designed so as to ease the study of other configurations by means of simple modifications of the free parameters.
The impact of noise and statistics on the obtained results should always be emphasized in a classroom discussion, to keep a practical approach.

A preliminary version of this project has been presented as part of the Quantum Computing course for the M.\,Sc. in Physics at the University of Ferrara, and received encouraging feedback. In particular, students mainly having  a theoretical background on quantum mechanics appreciate the possibility of visualizing those concepts in a practical simulation setting, and are thus encouraged in expanding their coding skills; whereas students who are already skilled in computer science appreciate the possibility of deepening their understanding of the underlying theoretical concepts. 

As a consequence, we find that this project has a significant educational impact. 
We hope that other students and professors will benefit from this tool, and will enjoy playing with the Bell-type inequalities.

\appendix
\vskip 1.cm

\section{Representing Bell-Wigner populations}
\label{app-rhombo}

As anticipated in sec.~\ref{sub-thfr-WB}, a particular configuration of the experimental setup allows for a captivating visualization, which may be useful at outreach events or in a classroom context. In eq. (\ref{eq-vectors}) let us consider the coplanar case with $\theta_{ab}=\pi/2$, $\phi=0$ and $\theta_{ac}=\pi/4$. Those three directions are orthogonal to three adjacent faces of the \textit{rhombicuboctahedron} (an Archimedes' solid), which can easily be realized using simple paper sheets, as shown in \cite{Wolf-rhombi}, or by using cheap snake puzzles.

\begin{figure}[h!]
\vskip .5cm 
 \begin{center}
\includegraphics[width=10cm]{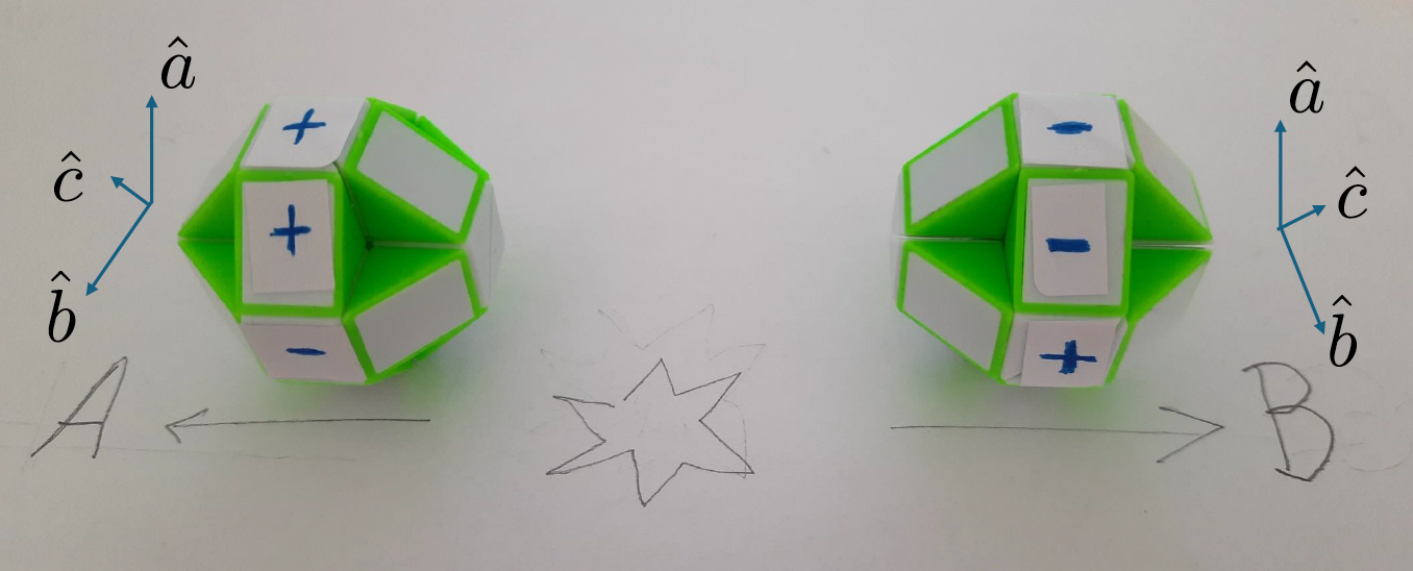}  
    \end{center}
\caption{\baselineskip=15 pt \small 
Visualization of the pair of particles belonging to population 3, flying apart and being measured by $A$ and $B$, along one of the coplanar directions $\hat a$, $\hat b$ and $\hat c$, with the predetermined result written on the faces orthogonal to the corresponding direction.}
\label{fig-pop3}
\vskip .5 cm
\end{figure}

We now take a rhombicuboctahedron as a representation of a single particle. According to Wigner's argument about LH variables, we choose three adjacent faces, corresponding to directions $\hat a$, $\hat c$ and $\hat b$ from top to bottom in fig.\,\ref{fig-pop3}, and label them with the predetermined values of the outcomes, as in tab.\,\ref{tab-pop} (where the ordering of the directions is $\hat a$, $\hat b$, $\hat c$). 
For population 1, for instance, the labels of particle 1 going towards Alice are $(\hat a +,\hat c  +, \hat b+)$ , those of particle 2 going towards Bob are $(\hat a -,\hat c-,\hat b-)$; 
for population 2 they are $(\hat a+,\hat c-,\hat b+)$ and $(\hat a-,\hat c+,\hat b-)$ for particles 1 and 2, respectively; 
for population 3, they are $(\hat a+,\hat c+,\hat b-)$ and $(\hat a-,\hat c-,\hat b+)$, as shown in fig. \ref{fig-pop3};
and so on, ending up with eight kinds of rhombicuboctahedrons. 

This visually helps to show how the LH variables have been introduced. The initial particles are assumed not to be all equal to each other when entangled in pairs: instead, we are pretending that they exist in eight different kinds, represented by the pre-labeled rhombicuboctahedrons, and the result of the possible measurements is already predetermined since the moment when the particles start to fly apart.

After the completion of this work, we became aware of a related visualization proposal, where cubes are used \cite{Levy:2024gnk}, the target being first-year undergraduate students or even high-school students.
While such representation may be useful to introduce the basic quantum formalism, our proposal allows for representing the violation of Bell-type inequalities for entangled particles, as the characterization of this condition requires non-orthogonal directions of the labels. A rhombicuboctahedron allows one to put labels representing spin eigenvalues along three coplanar directions separated by two angles of $\pi/4$: this is precisely the geometrical setup that maximally violates the Bell-Wigner inequality.

\section{Exact simulation of quantum computers in Qibo}
\label{app-qibo}

In classical simulation frameworks, quantum circuits can be simulated using Schrödinger's approach, which represents the quantum state of an $n$-qubit system by a complex vector $\ket{\psi}$ of dimension $2^n$. A quantum gate $G$ acting on a subset of $m<n$ qubits is represented by a unitary operator described by $2^m\times 2^m$
complex-valued matrix, and its application to the quantum state is performed through a matrix-vector multiplication as follows:
\begin{equation}
\psi'(\tau, q) = \sum_{\tau'} G(\tau, \tau') \psi(\tau', q),
\end{equation}
where $\tau$ and $q$ denote bitstrings indexing targeted and non-targeted qubits respectively, and the summation runs over all possible bitstrings $\tau'$.
For a concrete example, consider two qubits initialized in state $|00\rangle$, corresponding to the vector $\psi = (1,0,0,0)^T$. We are now going to apply two gates to this system, in particular an Hadamard gate and a NOT gate, respectively defined as:
\begin{equation}
    H = \frac{1}{\sqrt{2}}\begin{pmatrix}
    1 & 1 \\
    1 & -1 
    \end{pmatrix} 
    \qquad 
    \text{and}
    \qquad
   X = \begin{pmatrix}
    0 & 1 \\
    1 & 0 
    \end{pmatrix}.
    \label{eq:h_and_i}
\end{equation}
First, we apply the Hadamard gate to the first qubit:
\begin{equation}
\ket{\psi'} = (H \otimes I)\ket{\psi} = \frac{1}{\sqrt{2}}\begin{pmatrix}
1 & 0 & 1 & 0\\
0 & 1 & 0 & 1\\
1 & 0 & -1 & 0\\
0 & 1 & 0 & -1
\end{pmatrix}
\begin{pmatrix}1\\0\\0\\0\end{pmatrix} = \frac{1}{\sqrt{2}}\begin{pmatrix}1\\0\\1\\0\end{pmatrix},
\end{equation}
Where the unitary $( H \otimes I )$ is obtained through Kronecker product of the two matrices representing $H$ and $I$.
Then, we apply the NOT gate to the second qubit:
\begin{equation}
\ket{\psi''} = (I \otimes X)\,\ket{\psi'} = \begin{pmatrix}
0 & 1 & 0 & 0\\
1 & 0 & 0 & 0\\
0 & 0 & 0 & 1\\
0 & 0 & 1 & 0
\end{pmatrix}\frac{1}{\sqrt{2}}\begin{pmatrix}1\\0\\1\\0\end{pmatrix} = \frac{1}{\sqrt{2}}\begin{pmatrix}0\\1\\0\\1\end{pmatrix}.
\end{equation}
This example explicitly illustrates the standard Schrödinger simulation scheme using matrix-vector multiplications, which Qibo implements when executing a quantum circuit.



To give an example of how Qibo can be used to execute a quantum circuit, one can have a look at the following code:

\vspace{0.4cm}
\begin{lstlisting}[language=Python]
from qibo import Circuit, gates
from qibo import set_backend

# Set the most suitable backend for the execution
set_backend(backend="numpy")

# Let's prepare the |+> Bell's state
c = Circuit(nqubits=2)
c.add(gates.H(q=0))
c.add(gates.CNOT(q0=0, q1=1))
c.add(gates.M(0,1))

# Execute the circuit and collect output
outcome = c(nshots=1000)
\end{lstlisting}
\vspace{0.4cm}

In the previous code we simulate the circuit repeating the execution 1,000 times (\texttt{nshots}); the choice of the number of shots is particularly important because at each execution the system is first evolved to the final state and then measured in a chosen basis. This step forces us to make the state collapse on one of the possible occupied states, with a natural loss of information. Repeating the execution a suitable number of times allows us to reconstruct an approximation of the final state up to a statistical accuracy, which scales as $1/\sqrt{N_{\rm shots}}$ (see sec. \ref{sec-sn} for a more detailed discussion).

After the simulation, Qibo is able to return both the exact probabilities, as well as the measured frequencies\footnote{Frequencies can be computed in two ways with Qibo: a first lighter approach (which is used here) consists in sampling them from the probability distribution associated to the prepared final state. In practice, the circuit is executed once, and the frequencies are sampled from the prepared final state. A second, more intensive way, consists in simulating the collapse of the state \texttt{nshots} times, namely returning a list of collapsed states and not only the frequencies counter.}:
\vspace{0.4cm}
\begin{lstlisting}[language=Python]
# Exact probabilities 
probabilities = outcome.probabilities()

# Frequentist reconstruction
frequencies = outcome.frequencies()
\end{lstlisting}
\vspace{0.4cm}

The same code presented above can easily be run on any available classical multi-threading CPUs or GPUs, as well as on quantum hardware, by properly setting a dedicated backend among the ones offered by Qibo.

\section{Bell's theorem}
\label{app-Bell}

Following \cite{Bell:64}, we denote by $\lambda$ a hidden variable, that is a parameter that allows for a more complete specification with respect to quantum mechanics. The outcomes of the spin measurements thus depend on $\lambda$, and their correlation can be written as  
\beq
C(\hat a, \hat b) = \int d\lambda \rho(\lambda) \alpha(\hat a, \lambda) \beta(\hat b, \lambda)
\eeq
where $\rho(\lambda)$ is the probability distribution of $\lambda$ and is normalized to one.

Assuming the results of $A$ and $B$ along the same direction to be anti-correlated,
$\beta(\hat b, \lambda)=-\alpha(\hat b, \lambda)$,
and introducing the third direction $\hat c$, we have
\beq
C(\hat a, \hat b)-C(\hat a, \hat c) = - \int d\lambda \rho(\lambda) \left( \alpha(\hat a, \lambda) \alpha(\hat b, \lambda) -\alpha(\hat a, \lambda) \alpha(\hat c, \lambda) \right) \,\, .
\eeq
Exploiting the fact that $\alpha(\hat b, \lambda)^2=1$,
this can be written as
\beq
C(\hat a, \hat b)-C(\hat a, \hat c) 
=-\int d\lambda \rho(\lambda) \alpha(\hat a, \lambda) \alpha(\hat b, \lambda) 
\left(1- \alpha(\hat b, \lambda) \alpha(\hat c, \lambda)  \right) \,\,.
\eeq
Since $|\alpha(\hat a, \lambda) \alpha(\hat b, \lambda) |\leq 1$ while the last term in parenthesis in the integral above is non negative, we have 
\beq
|C(\hat a, \hat b)-C(\hat a, \hat c)| \leq
\int d\lambda \rho(\lambda) 
\left(1- \alpha(\hat b, \lambda) \alpha(\hat c, \lambda)  \right) = 1+ C(\hat b, \hat c) \,\,,
\eeq
where in the last step the normalization to one of $\rho(\lambda)$ was exploited.


\subsection*{List of abbreviations}
EPR, Eistein Podolski Rosen; CHSH, Clauser Horne Shimony Holt; LH, local hidden.

\section*{Declarations}

\subsection*{Availability of data and materials}
Data sharing is not applicable to this article: the data sets analysed during the current study are generated when the simulations are run. The code has been published in the \texttt{Qiboedu} GitHub repository, available under the GNU GPL 3 License at the short URL \href{https://cern.ch/bell-in-qibo}{https://cern.ch/bell-in-qibo}.

\subsection*{Competing interests}
The authors declare that they have no competing interests.

\subsection*{Authors' contributions}
IM conceived the project, and wrote the theoretical part. GLP and MR developed the simulation code and related figures. MG conceived the connection between theoretical results and the potential of quantum computing for didactic purposes, coordinating the various contributions. All authors contributed to the manuscript.

\subsection*{Acknowledgments}
IM thanks the CERN Theory Department for kind hospitality and support during the completion of this work. MR and MG are supported by CERN through the CERN Quantum Technology Initiative.
The authors thank Stefano Carrazza for useful discussions.


\bibliographystyle{elsarticle-num} 
\bibliography{bib} 
\end{document}